\documentclass[epj,nopacs]{svjour}
\usepackage{axodraw}
\usepackage{epsfig}
\usepackage{amsfonts}
\usepackage{amsmath}
\usepackage{bbm}
\allowdisplaybreaks[1]

\input pix.sty

\renewcommand{\eq}{eq.~}
\renewcommand{\eqs}{eqs.~}

\renewcommand{\fig}{fig.~}

\newcommand{\tinymsbar}{{\overline{\mbox{\tiny\rm{MS}}}}}

\newcommand{\Nf}{N_{\rm f}}

\newcommand{\Nc}{N_{\rm c}}

\newcommand{\mE}{m_\rmii{E}}
\newcommand{\gE}{g_\rmii{E}}
\newcommand{\gammaE}{\gamma_\rmii{E}}

\newcommand{\rmO}{{\mathcal{O}}}
\newcommand{\bmu}{\bar\mu}
\newcommand{\CA}{\Nc}

\def\lsi{\raise0.3ex\hbox{$<$\kern-0.75em\raise-1.1ex\hbox{$\sim$}}}
\def\gsi{\raise0.3ex\hbox{$>$\kern-0.75em\raise-1.1ex\hbox{$\sim$}}}
\newcommand{\lsim}{\mathop{\lsi}}
\newcommand{\gsim}{\mathop{\gsi}}

\newcommand{\rmii}[1]{{\mbox{\tiny\rm{#1}}}}

\newcommand{\Tint}[1]{{\hbox{$\sum$}\!\!\!\!\!\!\!\int\,}_{\!\!\!\!\raise-0.9ex\hbox{$\scriptstyle{#1}$}}}
\newcommand{\Tinti}[1]{{{\Sigma}\!\!\!\!\raise0.3ex\hbox{$\int$}_\rmii{${#1}$}}}

\newcommand{\bi}{\begin{itemize}}
\newcommand{\ei}{\end{itemize}}

\renewcommand{\tt}{\tilde t_0}

\newcommand{\hide}[1]{ }

\def\TAsc(#1,#2)(#3,#4,#5)%
{\SetWidth{2.0}\CArc(#1,#2)(#3,#4,#5)\SetWidth{1.0}}
\def\Lwidth{3}

\def\TAgl(#1,#2)(#3,#4,#5){\SetWidth{2.0}\PhotonArc(#1,#2)(#3,#4,#5){\Lwidth}%
{6.283 #3 mul 360 div #4 #5 sub #4 #5 sub mul sqrt mul Tdensity mul}%
\SetWidth{1.0}}
\def\TLgl(#1,#2)(#3,#4){\SetWidth{2.0}\Photon(#1,#2)(#3,#4){\Lwidth}
{#1 #3 sub #1 #3 sub mul #2 #4 sub #2 #4 sub mul add sqrt Tdensity mul}%
\SetWidth{1.0}}

\renewcommand{\picc}[1]{\;\parbox[c]{60pt}{\begin{picture}(60,30)(0,0)
\SetWidth{1.0}\SetScale{1.3} #1 \end{picture}}\; }
\def\Lwidth{1.3}

\begin{document}
\sloppy


\title{ A non-perturbative contribution to jet quenching }
\titlerunning{  A non-perturbative contribution to jet quenching }

\author{
M.~Laine
}

\institute{
Institute for Theoretical Physics, 
Albert Einstein Center, University of Bern, 
Sidlerstrasse 5, CH-3012 Bern, Switzerland}

\date{October 2012}
 
\abstract{%
It has been argued by Caron-Huot that infrared contributions
to the jet quenching parameter in hot QCD, denoted by $\hat q$, can 
be extracted from an analysis of a certain static-potential related 
observable within the dimensionally reduced effective field theory. 
Following this philosophy, 
the order of magnitude of a non-perturbative contribution 
to $\hat{q}$ from the colour-magnetic scale, $g^2 T/\pi$, is estimated. 
The result is small; it is probably below 
the parametrically perturbative but in practice slowly convergent
contributions from the colour-electric scale, whose 
all-orders resummation therefore remains an important challenge.
}

\maketitle


%
\section{Introduction}
\la{se:intro}

One of the classic qualitative indications for the generation
of a thermal medium in heavy ion collision experiments 
is the fact that high-$p_T$ jets get quenched~\cite{exp1,exp2}. 
In fact, not only do 
jets get quenched but they appear to do so more effectively than  
naive estimates suggest~\cite{phen1,phen2}. This motivates not only 
a complete leading-order weak-coupling analysis~\cite{ax},
but also developing methods to go beyond the leading order~\cite{sch}, taking 
into account large corrections from the infrared (IR) scales 
that are characteristic of thermal field theory~\cite{linde,gpy}.
(The weak-coupling regime has recently also 
been discussed in ref.~\cite{other}.)

In a weakly coupled non-Abelian plasma, 
there are two momentum scales in the IR: the colour-magnetic 
scale, $g^2 T/\pi$, and the colour-electric scale, $gT$
($g^2 = 4\pi \alpha_s$ is the QCD gauge coupling).
We refer to these through the gauge coupling, $\gE^2/\pi$, 
and the mass parameter, $\mE^{ }$, of the dimensionally reduced
effective theory~\cite{dr1,dr2}. 
The contributions of the two scales have been disentangled 
for several thermodynamic observables, and a particularly 
rich source of information is the ``spectrum'' measured through
screening masses~\cite{ay}. For instance, the smallest 
screening mass (at temperatures below $\sim 10$ GeV), 
which is parametrically of the form 
$2\mE^{ } + \rmO(\gE^2/\pi)$, gets a numerically insignificant 
contribution from the colour-magnetic scale~\cite{lv}, 
whereas the Debye screening mass, which is parametrically
of the form $\mE^{ } + \rmO(\gE^2/\pi)$, is in practice all but 
dominated by the colour-magnetic scale~\cite{lp2}: 
\ba
 M_\rmii{D} & = & \mE^{ } + \frac{\gE^2 \CA }{4\pi}
 \biggl(\ln\frac{\mE^{ }}{\gE^2} + 6.9 \biggr) + 
 \rmO( g^3 T )
 \nn & \simeq & 
 \mE^{ } + 1.65 \,\gE^2+ \rmO(g^3 T)
 \;.
\ea  
The purpose of the current study is to estimate whether
one of these extreme scenarios could be relevant for $\hat{q}$.

%
\section{Colour-electric contribution}

The rate of jet broadening 
(transverse momentum diffusion)
is often parametrized by a phenomenological
coefficient, called the jet quenching parameter 
and denoted by $\hat{q}$~\cite{qhat_def}.
Although, strictly speaking, the definition of $\hat{q}$ is ambiguous
even at leading order~\cite{bdmps}, it can be argued that the ambiguities 
can be hidden by considering a quantity called the transverse
collision kernel, $C(q^{ }_\perp)$; then 
\be
 \hat{q} = \int_0^{q^*} 
 \! \frac{{\rm d}^2 \vec{q}^{ }_\perp}{(2\pi)^2} \, 
 q_\perp^2 C(q^{ }_\perp)
 \;. \la{hq_full}
\ee
It is now the choice of the upper bound $q^*$ which reflects 
the ambiguities. The form of $C(q^{ }_\perp)$ for $q^{ }_\perp\sim \pi T$ was 
determined in ref.~\cite{ax} and for $q^{ }_\perp \sim gT$
in ref.~\cite{sch}; although the NLO contribution from $q^{ }_\perp \sim gT$
to $\hat{q}$
is parametrically suppressed by $\rmO(g)$, it is numerically large. 

More precisely, denoting by $\mE^{ } = g T \sqrt{\Nc/3 + \Nf/6} + \rmO(g^3T)$ 
the Debye mass parameter
and by $\gE^2 = g^2 T + \rmO(g^4T)$ the coupling constant of the dimensionally 
reduced effective field theory, the next-to-leading order (NLO)
expression for the IR part ($q^{ }_\perp \sim \mE^{ }$)
of the collision kernel can be written as \cite{sch} 
\ba
 C(q^{ }_\perp) & = & {\gE^2 C_F} \biggl\{ 
  \frac{1}{q_\perp^2} - \frac{1}{q_\perp^2 + \mE^2}
 \biggr\} 
 \nn  & + & {\gE^4 C_F \CA}
 \biggl\{ 
 \frac{7}{32 q_\perp^3} 
   -  \frac{\mE^{ } + 2 (q^{ }_\perp- \frac{\mE^2}{q^{ }_\perp} ) 
   \arctan(\frac{q^{ }_\perp}{\mE^{ }} )}
   {4 \pi (q_\perp^2+\mE^2 )^2}
 \nn & & \; 
 +\,  \frac{\mE^{ }}{4\pi (q_\perp^2 + \mE^2)}
 \biggl[ \frac{3}{q_\perp^2 + 4 \mE^2} - \frac{2}{q_\perp^2 + \mE^2}
 - \frac{1}{q_\perp^2}
 \biggr]
 \nn & & \; 
 -\, \frac{\arctan(\frac{q^{ }_\perp}{\mE^{ }} )}
 {2 \pi q^{ }_\perp (q_\perp^2+\mE^2 )}
 + \frac{\arctan(\frac{q^{ }_\perp}{2 \mE^{ }} )}{2 \pi q_\perp^3} 
  \la{kernel} \\ & & \; 
 +\,  \frac{\mE^{ } - (\frac{q^{ }_\perp}{2} +  \frac{2 \mE^2}{q^{ }_\perp} ) 
   \arctan(\frac{q^{ }_\perp}{2 \mE^{ }} )}
   {8 \pi q_\perp^4}
 \biggr\} 
 + \rmO(\gE^6)
 \;. \nonumber
\ea
Here $C_F = (\Nc^2 - 1)/(2 \Nc)$ is the Casimir of the fundamental 
representation. 
Choosing $\mE^{ } \ll q^* \ll \pi T$, 
the integral in \eq\nr{hq_full} yields
\ba
 \hat{q} & = &  
  \frac{\gE^2 \mE^2 C_F}{2\pi} \ln \frac{q^*}{\mE^{ }}
 +  \frac{\gE^4 \mE^{ } C_F \CA}{2\pi}
 \biggl\{ 
   - \frac{q^*}{16\mE^{ }} 
 \nn & & \quad + \, 
  \frac{3\pi^2 + 10 - 4 \ln 2}{16\pi } + 
 \rmO\Bigl( \frac{\mE^{ }}{q^*} \Bigr)
 \biggr\}
 + \rmO(\gE^6) 
 \;. \hspace*{1cm} \la{mE_contr}
\ea
The terms involving $q^*$ cancel against 
contributions from hard momenta, 
$q^{ }_\perp \gsim \pi T$~\cite{ax}, and 
the $q^*$-independent middle term inside
the curly brackets then represents the physical 
NLO contribution to $\hat{q}$~\cite{sch}. 

%
\section{Colour-magnetic contribution}

For small momenta, $q_\perp^{ } \ll \mE^{ }$, the NLO
part of \eq\nr{kernel} behaves  as 
\be
 C^\rmi{NLO}(q^{ }_\perp) = \gE^4 C_F \CA 
 \biggl\{ 
  \frac{7}{32 q_\perp^3} 
 - \frac{1}{48 \pi \mE^{ } q_\perp^2}
 - ... 
 \biggr\} 
 \;. \la{small}
\ee
We note from \eqs\nr{kernel}, \nr{small} that, as also suggested
by the operator definition~\cite{sch}, the 
small-$q^{ }_\perp$ limit ($q^{ }_\perp \ll \mE^{ }$) 
of $C(q^{ }_\perp)$ goes over
into minus the momentum-space static potential of three-dimensional 
pure Yang-Mills theory:  
\be
 \lim_{q^{ }_\perp \ll\; \mE } \;
 C(q^{ }_\perp) = - \tilde V(q^{ }_\perp)
 \;, \la{id}
\ee
where~\cite{ys_thesis}
\be
 \tilde V(q^{ }_\perp) = 
 - \frac{\gE^2 C_F}{q_\perp^2} 
 - \frac{7 \gE^4 C_F \CA}{32 q_\perp^3} + 
 \rmO\Bigl( \frac{\gE^6}{q_\perp^4} \Bigr)
 \;. \la{q_pert}
\ee
(In principle higher-dimensional ``hybrid'' potentials 
could also appear, but their contributions to $\hat{q}$ are 
suppressed by $(\gE^2/\pi\mE^{ })^n$, with some positive $n$, with respect 
to those from $\tilde V(q^{ }_\perp)$.) 

However, at the next order the perturbative 
expansion breaks down: a direct momentum-space 
computation of $\tilde{V}(q^{ }_\perp)$  at 2-loop order 
produces in $d=3-2\epsilon$ 
dimensions the (gauge-independent) expression~\cite{ys_thesis}
\be
 \tilde{V}^\rmi{NNLO}(q^{ }_\perp) = 
 \frac{\gE^6 C_F \CA^2}{(4\pi)^{1-2\epsilon} q_\perp^{4+4\epsilon}}
 \biggl(
  \frac{1}{4\epsilon^2} + \frac{3}{4\epsilon} 
 \biggr)
 \;. \la{IR}
\ee
A comparison with \eq\nr{hq_full} suggests the presence of a 
non-perturbative contribution to $\hat{q}$ at $\rmO(\gE^6)$, both
because the integral is logarithmically divergent at the lower end,
and because the coefficient in \eq\nr{IR} is IR divergent. 

For future reference we note that 
the Fourier transform of \eq\nr{q_pert},  
$V(r) = \int_{\vec{q}^{ }_\perp}^{ } 
( e^{i \vec{q}^{ }_\perp \cdot \vec{r}} - 1) \tilde{V}(q^{ }_\perp)$, 
yields the coordinate space potential
\be
 V(r) = \frac{\gE^2 C_F}{2\pi} \, \ln\frac{r}{r_*} + 
 \frac{7 \gE^4 C_F \CA }{64\pi } \, r 
 + \rmO(\gE^6 r^2)
 \;, \la{r_pert}
\ee
where $r_*$ is a regularization-dependent constant. 

%
\section{Transformation to configuration space}

In view the delicate nature of the IR 
contribution to $\hat{q}$, 
it is useful to re-express the small-$q^{ }_\perp$ 
part of \eq\nr{hq_full} in another way. The idea 
is to make use of \eq\nr{id}, in combination with a non-perturbative
understanding of $V(r)$, in order to get a handle on IR physics. 

To be concrete, 
let $\tilde{\theta}^{ }_\rmii{$\!\Lambda$}(q^{ }_\perp)$ 
be some cutoff function, 
where $\Lambda$ is chosen to be formally
in the range $\gE^2/\pi  \ll \Lambda \ll \mE^{ } $. 
Then we can rephrase the IR part of 
\eq\nr{hq_full}, to be denoted by $\hat{q}^{ }_\rmii{$\Lambda$}$, as  
\be
 \hat{q}^{ }_\rmii{$\Lambda$} =  
 \int \! \frac{{\rm d}^2 \vec{q}^{ }_\perp}{(2\pi)^2}
 \, q_\perp^2 \, C(q^{ }_\perp) 
 \, \tilde{\theta}^{ }_\rmii{$\!\Lambda$}(q^{ }_\perp) 
 = 
 \int \! {\rm d}^2 \vec{r}  \, \nabla^2  V(r) \, 
 \theta^{ }_\rmii{$\!\Lambda$}(r) 
 \;, \la{qhat_ir}
\ee
where 
$
 \theta^{ }_\rmii{$\!\Lambda$}(r) \equiv 
  \int \! \frac{{\rm d}^2 \vec{q}^{ }_\perp}
 {(2\pi)^2} e^{- i \vec{q}^{ }_\perp \cdot\vec{r}} \, 
  \tilde{\theta}^{ }_\rmii{$\!\Lambda$}(q^{ }_\perp)
$
is the configuration space version of the cutoff function. 
For instance, if $\tilde{\theta}^{ }_\rmii{$\!\Lambda$}(q^{ }_\perp)
 = \theta(\Lambda - q^{ }_\perp)$, 
then $\theta^{ }_\rmii{$\!\Lambda$}(r) = \Lambda\, J_1(\Lambda r) / 2 \pi r$. 
This very choice is not particularly convenient, however, since the Bessel 
function $J_1$ is oscillatory. 
We find it more transparent 
to choose a Gaussian, 
\be
 \tilde{\theta}^{ }_\rmii{$\!\Lambda$}(q^{ }_\perp) \equiv 
 \exp\Bigl( - \frac{q_\perp^2}{\Lambda^2} \Bigr)
 \;, \quad
 \theta^{ }_\rmii{$\!\Lambda$}(r) = \frac{\Lambda^2}{4\pi} 
 \exp\Bigl( - \frac{r^2 \Lambda^2 }{4} \Bigr) 
 \;,
\ee
because both functions are elementary and positive. 

Now, 
because of rotational symmetry and the form of \eq\nr{r_pert}, 
the behaviour of $\nabla^2 V(r)$ 
appearing in \eq\nr{qhat_ir} is 
\ba 
 \nabla^2 V(r) & = &  \gE^2 C_F \, \delta^{(2)}(\vec{r}) + 
 \biggl[ V''(r) + \frac{V'(r)}{r} \biggr]_{r>0}
 \nn & = & 
 \gE^2 C_F \, \delta^{(2)}(\vec{r}) + 
 \biggl[ \frac{2 c(r)}{r^3} + \frac{F(r)}{r} \biggr]_{r>0}
 \;,   
\ea
where we have adopted the notation $F(r) \equiv V'(r)$ and 
$c(r) \equiv r^3 V''(r) / 2$ from ref.~\cite{lw}.

We subsequently 
need to subtract the known perturbative terms, \eq\nr{r_pert}, from
the IR contribution to $\hat{q}^{ }_\rmii{$\Lambda$}$, given that they
were already included in \eqs\nr{kernel}, \nr{mE_contr}.
The corresponding potential, i.e.\ the term of 
$\rmO(\gE^6 r^2)$ in \eq\nr{r_pert}, is denoted by $\delta V$.
Thereby we obtain an expression for the remaining IR 
contribution, let us call it $\delta \hat{q}^{ }_\rmii{$\Lambda$}$:
\ba
 \delta \hat{q}^{ }_\rmii{$\Lambda$} 
 & = &  
 2 \pi 
 \int_{0^+}^\infty \! {\rm d} r  \, 
 \biggl[ F(r) + \frac{2 c(r)}{r^2} - \frac{7 \gE^4 C_F \CA}{64\pi}
 \biggr] \, \theta^{ }_\rmii{$\!\Lambda$}(r) 
 \nn 
 & = &
 \fr12  
 \int_{0^+}^\infty \! {\rm d} \hat{r}  \; 
 \Lambda\, \phi \Bigl( \frac{\hat{r}}{\Lambda} \Bigr) \; 
 \exp\Bigl(-\frac{\hat{r}^2}{4}\Bigr)
 \;, \la{del_qhat}
\ea
where we rescaled the integration variable as $\hat{r} = r \Lambda$ 
and defined
\ba
 \phi(r)\; & \equiv & \; 
 \delta V'(r) + r \delta V''(r) \la{phi_def}
 \\ \; & = & \; 
 F(r) + \frac{2 c(r)}{r^2} - \frac{7 \gE^4 C_F \CA}{64\pi}
 \;. 
\ea
As \eq\nr{del_qhat} shows, only the short-distance part of $\phi(r)$, 
terms linear in $r$ (modulo logarithms), contributes to 
$\delta \hat{q}^{ }_\rmii{$\Lambda$}$; 
this corresponds to terms quadratic in $r$ in $\delta V(r)$.

The nature of the short-distance behaviour of $\delta V(r)$ can be discussed 
within a framework similar to the Operator Product Expansion~\cite{mv,hl}. 
The lowest-dimensional condensate in three-dimensional
pure SU(3) evaluates to 
\ba
 & & \hspace*{-1cm}
 \frac{1}{2 \gE^2} \Bigl\langle
 \tr [F_{kl} F_{kl}] 
 \Bigr\rangle_\tinymsbar 
  \;=\;   
 \frac{ 6 \gE^6 C_F \CA^4}{(4\pi)^4} 
 \nn & & \; \times \,  \biggl[ 
  \biggl( \frac{43}{12} - \frac{157}{768} \pi^2 \biggr) 
 \biggl( 
 \ln\frac{\bmu}{2 \CA \gE^2} -\fr13
 \biggr)
 \nn & & \quad - \, 0.2 \pm 0.4^\rmii{(MC)} \pm 0.4^\rmii{(NSPT)}
 \biggr]
  \;, \la{condensate}
\ea
where the errors come from Monte Carlo simulations (MC)~\cite{lattg6} and 
a scheme conversion (NSPT)~\cite{nspt_mass}, and the numerical values
apply for $\Nc = 3$. Just cancelling the scale dependence from 
\eq\nr{condensate} one might expect a short-distance behaviour of the form  
\be
 \delta V(r) \sim \gE^6 r^2 \ln \Bigl(\frac{1}{\gE^2 r }\Bigr)
 + \rmO(\gE^8 r^3)
 \la{naiveV}
 \;.
\ee
However, according to ref.~\cite{ps} the 
shape at $\gE^2 r \ll 1$ is really more complicated, 
$\sim {\gE^6  r^2 \ln [\ln (1/\gE^2 r)]} / {\ln (1 / \gE^2 r)}$,   
where $\gE^2\Nc\ln (1/\gE^2 r)$ represents the difference of octet and 
singlet potentials. Note that such a tail does not contribute in 
\eq\nr{del_qhat} for $\Lambda  \gg \gE^2/\pi$.
Unfortunately 
the \linebreak analysis is not valid for the range $\gE^2 r  \sim 1$
that is relevant for us here, so we resort to modelling in the following. 

%
\section{Modelling of lattice data}

In \fig\ref{fig:qhat}, numerical data for $r_0^2 \phi(r)$ from 
ref.~\cite{lw} is shown. The parameter $r_0$ is defined from 
\be
 r_0^2 F(r_0) = 1.65 \la{r0}
\ee
for any given lattice spacing~\cite{rs}. 
In the continuum limit~\cite{lw}, 
\be 
 \gE^2 r_0 \approx 2.2
 \;, \la{r02}
\ee 
and we have inserted this estimate in order to combine 
the last term of \eq\nr{phi_def} with lattice data. 
(Obviously, it would be nice to add data at shorter distances, 
but this task is non-trivial 
because strong cutoff effects appear and a careful extrapolation to the 
continuum limit is needed.) 

\begin{figure}[t]


\centerline{%
\epsfysize=8.0cm\epsfbox{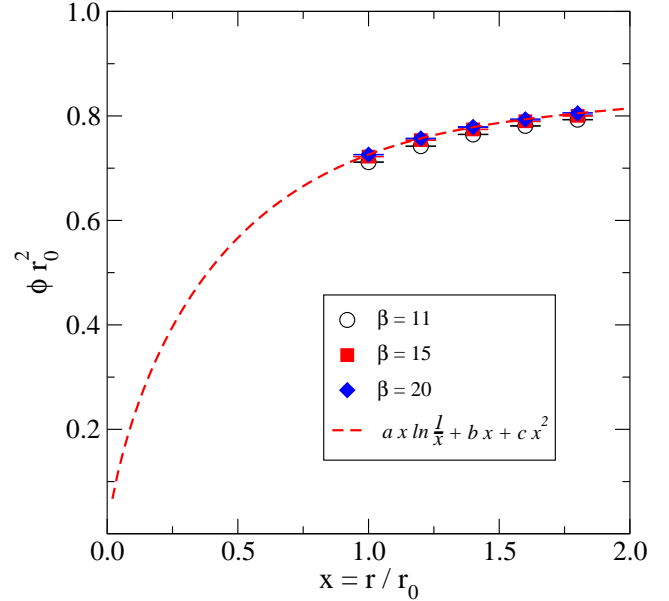}%
}


\caption[a]{\small
 The function $\phi$ defined by \eq\nr{phi_def}, 
 in units of the parameter $r_0$, defined by \eq\nr{r0}. 
 The lattice data is from tables 5 and 7 of ref.~\cite{lw}.
}

\la{fig:qhat}
\end{figure}

Now, motivated by the naive \eq\nr{naiveV} and the definition
in \eq\nr{phi_def}, 
we describe the data through the function
\be 
 r_0^2 \phi(r) =  \frac{a\, r}{r_0}  
 \ln  \frac{r_0}{r} + 
 \frac{b\, r}{r_0}  +   \frac{c \, r^2}{r_0^2} 
 \;, \quad r \lsim 2 r_0 \;. \la{model}
\ee
The data (at $\beta = 20$) are well modelled 
($\chi^2 / $d.o.f.\,$ = 0.18$) 
with $a=0.72(2)$, $b=0.55(1)$, $c=0.18(1)$. 
We remark that around $r/r_0 \approx 2.0$ this function is  
close to the asymptotic value  given by the 
non-perturbative string tension minus the perturbative subtraction,  
$\{ [0.553(1)]^2 - 7/16\pi \} \gE^4 r_0^2 \approx 0.81$~\cite{mt}. 
We have checked that adding a cubic term to the ansatz of \eq\nr{model} 
does not change the results in any significant way, 
for instance $a=0.74(14)$, $b=0.52(17)$ then.
In the following it is therefore assumed that \eq\nr{model} 
reflects the magnitude of the true function for $ \gE^2 r \sim 1$. 

Carrying out the integral in \eq\nr{del_qhat}, we get 
\be
 \delta \hat{q}^{ }_\rmii{$\Lambda$} 
 = 
 \frac{1}{r_0^3}
 \biggl[
   a \biggl(
    \ln\frac{\Lambda r_0}{2} + \frac{\gammaE}{2} 
   \biggr) 
   + b + \rmO\Bigl( \frac{1}{\Lambda r_0} \Bigr)
 \biggr]
 \;.   \la{delta_hat}
\ee 
The coefficient $a/r_0^3$, which comes together
with a logarithmic dependence on the cutoff $\Lambda$, 
should be an analytically computable function, since the cutoff 
should cancel against 
an NNLO contribution from the colour-electric scale;
in this sense it is not a ``genuine'' colour-magnetic contribution.  
In contrast, 
the coefficient $b/r_0^3$ represents
a genuine colour-magnetic contribution 
{\em within the model}  
(although, because of the freedom in choosing the argument 
of the logarithm, this notion is somewhat ambiguous). 

%
\section{Phenomenological interpretation} 

Omitting from \eq\nr{delta_hat}  
terms vanishing for $\Lambda r_0 \gg 1$; replacing
the cutoff inside the logarithm with the scale $\sim \mE^{ }$ at which 
other physics sets in; inserting $r_0$ from 
\eq\nr{r02}; and estimating $\mE^{ }/\gE^2 \sim 1$ 
(cf.\ fig.~1 of ref.~\cite{lv}); we get 
the order-of-magnitude estimate 
\be
 \delta \hat{q}^{ }_\rmii{$\Lambda$} 
 \simeq \frac{\gE^6}{2.2^3}
 \Bigl[ 0.72 
   \biggl(
    \ln\frac{2.2}{2} + \frac{\gammaE}{2} 
   \biggr) 
   + 0.55
 \Bigr]
 \; \approx \; 0.08 \, \gE^6
 \;. \la{num_gE}
\ee
This can be compared with the middle term from
the curly brackets in \eq\nr{mE_contr}, 
\be
 \delta \hat{q} \approx \frac{\gE^4 \mE^{} C_F \CA}{32\pi^2}
 (3\pi^2 + 10 - 4\ln 2)
 \; \approx \; 0.47 \, \gE^4 \mE^{ } 
 \;. \la{num_mE}
\ee
The colour-magnetic contribution, \eq\nr{num_gE}, is clearly
below the NLO perturbative
contribution from the colour-electric scale, \eq\nr{num_mE}. 

In conclusion, the contribution to $\hat{q}$ from the colour-magnetic
scale, $q^{ }_\perp \sim \gE^2/\pi$,
may well be of modest magnitude. Thus the 
phenomenological motivation for its theoretically
consistent determination may be feeble; rather, 
it should probably be measured as  a part of the total IR contribution from 
both the colour-electric and colour-magnetic scales, e.g.\ along the 
lines suggested in ref.~\cite{sch}. 

%
\section*{Acknowledgements}

This work was partly supported by the Swiss National Science Foundation
(SNF) under grant 200021-140234.

%
\section*{Note added}

Recently a paper appeared~\cite{new} in which, as acknowledged there, 
some basic ideas of the current study, communicated to one of the authors
several years ago, were revealed. Unfortunately
the practical implementation, citing e.g.\ a peculiar temperature
dependence of $\delta \hat{q}^{ }_\rmii{$\Lambda$}$, 
appears to suffer from misunderstandings. (After the appearance of 
the current paper and further email correspondence, 
many of these problems have been rectified in v2.)

\appendix
\renewcommand{\thesection}{Appendix~\Alph{section}}
\renewcommand{\thesubsection}{\Alph{section}.\arabic{subsection}}
\renewcommand{\theequation}{\Alph{section}.\arabic{equation}}


\end{document}